# AN INVESTIGATION OF HERTZIAN CONTACT IN SOFT MATERIALS USING PHOTOELASTIC TOMOGRAPHY

By

Benjamin Mitchell[1], Yuto Yokoyama[2], Ali Nassiri[3], Yoshiyuki Tagawa[2], Yannis P. Korkolis[3] and Brad L. Kinsey[1*]

*Corresponding author. e-mail: brad.kinsey@unh.edu

[1]Department of Mechanical Engineering, University of New Hampshire,

Durham, NH 03824, USA

[2]Department of Mechanical Systems Engineering, Tokyo University of Agriculture and

Technology, Tokyo, 184-8588, Japan

[3]Department of Integrated Systems Engineering, The Ohio State University

1971 Neil Avenue, Columbus, OH 43210, USA

## Abstract

Hertzian contact of a rigid sphere and a highly deformable soft solid is investigated using integrated photoelasticity. The experiments are performed by pressing a styrene sphere of 15 mm diameter against a 44 x 44 x 47 $mm^3$ cuboid made of 5% wt. gelatin, inside a circular polariscope, and with a range of forces. The emerging light rays are processed by considering that the retardation of each ray carries the cumulative effect of traversing the contact-induced axisymmetric stress field. Then, assuming Hertz's theory is valid, the retardation is analytically calculated for each ray and compared to the experimental one. Furthermore, a finite element model of the process introduces the effect of finite displacements and strains. Beyond the qualitative comparison of the retardation fields, the experimental, theoretical, and numerical results are quantitatively compared in terms of the maximum equivalent stress, surface displacement, and contact radius dimensions. A favorable agreement is found at lower force levels, where the assumptions of Hertz theory hold, whereas deviations are observed at higher force levels. A major discovery of this work is that at the maximum equivalent stress location, all three components of principal stress can be determined experimentally, and show satisfactory agreement with theoretical and numerical ones in our measurement range. This provides valuable insight into Hertzian contact problems since the maximum equivalent stress controls the initiation of plastic deformation or failure. The measured displacement and contact radii also reasonably agree with the theoretical and numerical ones. Finally, the limitations that arise due to the linearization of this problem are explored.

# 1. Introduction

The contact of a sphere with the flat surface of a homogeneous material presents a useful platform for analyzing the mechanics of Hertzian contact. The stress field within the material is of particular importance because it can be used to identify the contact zone and the region where the onset of yielding, or failure, occurs. The pioneering work of Huber (1904) showed that the point of maximum equivalent stress occurs inside the material, beneath the surface, and is the region where plastic deformation is initiated [1]. Hence, the magnitude of stress at this crucial point is highly sought-after. Analytical and numerical models have been used to calculate the stress inside a material subject to Hertzian sphere loading [2-6]; however, experimental methods are far less prevalent. This is due to the difficulty in measuring stresses inside of a material subject to mechanical deformation. One tool that is particularly suited for this is photoelasticity.

Photoelasticity is a stress analysis technique that correlates polarized light with the principal stress difference. A light ray traveling through a stressed, birefringent solid incrementally acquires a change in phase, resulting in a cumulative phase retardation (denoted $\delta$) after it has emerged from the photoelastic body. The technique gained widespread adoption for decades, due to its ability to analyze complex stress fields, typically, however, under plane-stress. In these cases a direct relationship between the principal stress difference and the measured optical phase retardation exists [14, 15]. The method remains relevant to this date due its nondestructive nature, whole-field graphic capability, visual appeal, and relative ease of testing. It has shown success in measuring stress distributions in a variety of mechanical testing specimens, e.g., residual stress in glass, and the determination of stress concentration factors [7-13]. Despite these achievements, there are several restrictions to the method. Applications in 3D stress states have always remained limited, due to challenges in both performing and interpreting the experiments. Circumstances where the stress state varies along the light propagation direction do not admit trivial solutions. This is due to the complex propagation qualities of light passing through optically anisotropic media, which render the governing equations non-linear and make the problem ill-posed [16-19]. Despite these constraints, there has been success in analyzing 3D stress states using the approximation of geometrical optics in integrated photoelasticity [20-22]. In that sense, integrated photoelasticity can be a useful tool for analyzing the stress field in 3D Hertzian contact problems.

The aim of this study is to analyze the extent that classical contact mechanics can be used to study Hertzian contact with a soft, highly-deformable, elastic substrate, which has numerous uses in the medical, soft robotics, pharmaceutical, and culinary fields. For this, the sub-surface stress fields are probed using integrated photoelasticity and are compared to the predictions of the classical theory. Through this, information can be gathered about the contact zone, and the maximum equivalent stress can be quantified. The experimental phase retardation fields are compared with the theoretical ones, calculated using Hertzian contact theory. Analysis of a soft material subject to Hertzian contact through photoelasticity is an innovation on this classical mechanics problem and is important because it serves as a tool for nondestructive stress measurement, while providing verification of the approximation of geometrical optics and Hertzian contact models applied to soft solids. Complementary to this work, the orientation of principal stress of Hertzian contact is investigated in a companion paper [23]. This comparison allows for verification of the integrated photoelasticity model and is also used to determine its limitations. The results and methodologies used in these studies also serve as a benchmark for future studies, such as the impact of a droplet on the soft solid.

This paper is organized as follows. Section 2 describes the experimental methods (for more details, see the companion paper [23]), while Section 3 derives the theoretical stresses expected in the Hertzian loading scenario. These stresses are used in the optically equivalent model to obtain the theoretical phase retardation fields, which can be directly compared to the experiments. In addition to theoretical stresses, numerical determination of stresses induced in the Hertzian contact problem are simulated using finite elements which, unlike the classical theory, use the finite strain formulation and allow for large displacements to occur. This is described in Section 4 where simulations are used in conjunction with the integrated photoelasticity model and compared with theory and experiments. Section 5 discusses the results and demonstrates the ability to determine equivalent stress and each principal stress component at the point of highest stress, given only the phase difference field obtained from integrated photoelasticity. This is a major contribution of this work. Section 6 summarizes the findings and provides recommendations for future studies. Future research will study the ability to measure equivalent stress and principal stress components for dynamic water droplet impact on gelatin media [24, 25]. This will help to understand further the erosion mechanisms observed during droplet impact on a substrate, e.g., in the novel material removal process termed Water Droplet Machining (WDM) [26, 27].

## 2. Experimental methodologies

To probe the stress state inside of a material subject to a Hertzian contact, an experiment is devised using integrated photoelasticity. A schematic of the setup is shown in Fig. 1a where a 15 mm diameter styrene sphere is pressed against the top surface of a 44 x 44 x 47 mm$^3$ gelatin cuboid. This produces a Hertzian contact scenario in the vicinity of the sphere, establishing an axisymmetric stress field within the gelatin. Polarized light is incident upon one of the faces of the cuboid. Since gelatin is birefringent, as the light propagates through the cuboid, it accumulates phase retardations corresponding to the state of stress along each light ray. The light rays then emerge from the back side of the cuboid and are acquired for analysis. An (x, y, z) Cartesian coordinate system is chosen for the cuboid where the y-direction coincides with the light propagation direction, see Fig. 1a. Since the contact-induced stress field in the cuboid exhibits axial symmetry, a cylindrical coordinate system is also adopted as in Fig. 1a. The r-z plane of the cylindrical coordinate system is shown in Fig. 1b, where axial symmetry is assumed. The use of this coordinate system will be further elaborated in Section 4.

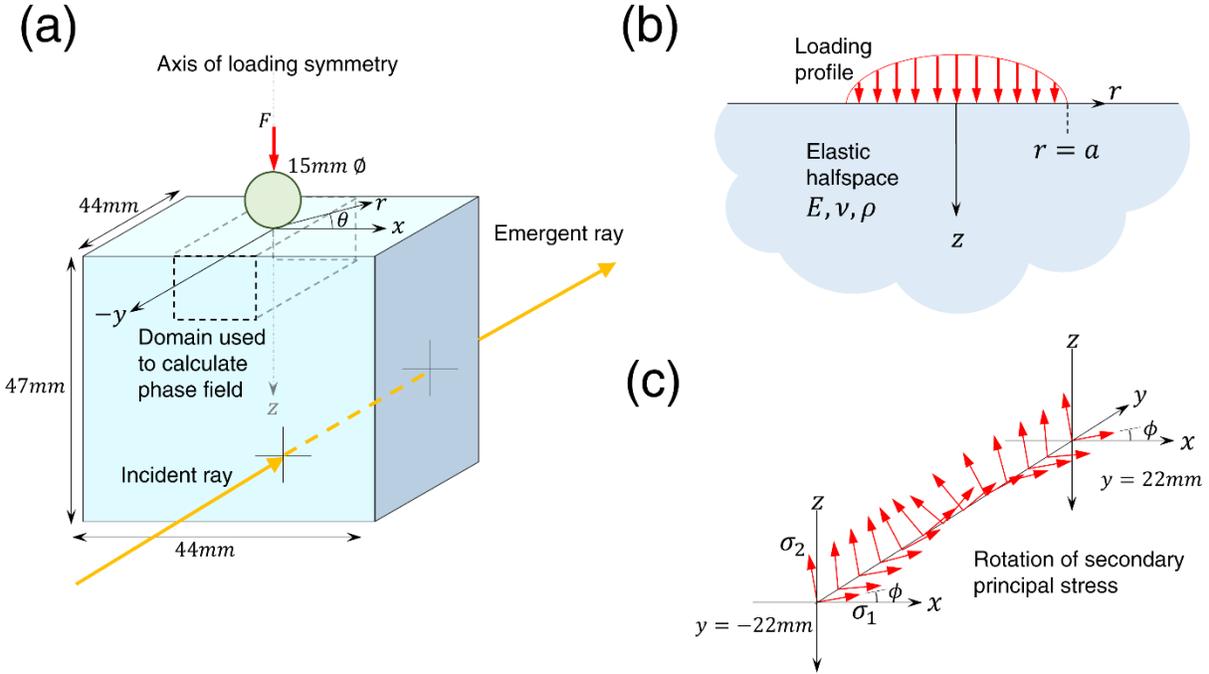

Figure 1: (a) Schematic of gelatin cuboid subject to axisymmetric Hertzian contact at center of top surface. Circular polarized light is incident upon the front surface of the cuboid (y = -22 mm), with emergent light exiting the back side of the cuboid (y = 22 mm), including accumulated stress-induced phase retardation $\delta$. (b) Diagram of axisymmetric Hertzian loading configuration on top surface of an elastic half-space. (c) Illustration of secondary principal stress rotation along a given light ray.

## 2.1 Material preparation and properties

The gelatin is created by mixing gelatin powder from porcine skin (Sigma Aldrich G6144-1KG) with hot water (90°C) at a concentration of 5% wt. The solution is stirred until the temperature reaches 30 °C; then the mixture is poured into an acrylic container with inside dimensions of 44 x 44 x 47 mm$^3$ . The filled container is then placed in a refrigerator at 4 °C to allow solidification and left at that temperature for at least 18 hrs. Before conducting the experiments, the container is taken out of the refrigerator and given 2 hrs to reach room temperature (20°C).

It is well-established that gelatin can be considered as a linearly-elastic material, even to very large strains. The Young's Modulus of the gelatin is determined using the surface deformation technique [28, 29]. Here, a sphere is pressed against the centroid of the top surface of the gelatin with force $F$, and a camera is used to determine the maximum surface displacement induced by the sphere, which occurs directly underneath the sphere along the z-axis. This is repeated for a range of applied forces. The relationship between maximum surface displacement and the force the sphere applies to the gelatin is given by [6, 30]:

$$u_z^{max} = \left(\frac{9F^2}{8DE^{*2}}\right)^{1/3}, \quad (1)$$

where $D$ is the diameter of the sphere, and $E^*$ is the Effective Modulus of the material, given by:

$$E^* = \frac{E}{1-v^2}, \quad (2)$$

where $v$ is the Poisson Ratio. Due to the nearly incompressible properties of gelatin at small strains, its Poisson Ratio is taken in this study as $v$ = 0.49 [28, 31, 32]. Using Eqs. (1) and (2), along with the experimentally determined maximum displacements, the Elastic Modulus is calculated to be 4.22 kPa. This value has reasonable agreement with other established Moduli of gelatin [31-33]. The density of gelatin at 5% wt. is 1,010 kg/m³ [34]. The Stress-Optic Coefficient of gelatin, which is a material property, has been determined to be $C$ = 3.3 x $10^{-8}$ Pa$^{-1}$ by fitting the maximum theoretical to maximum experimental phase retardation [23]. This value is in good agreement with previously established values for gelatin [8, 35]. The material properties and experimental parameters used in the study are summarized in Table 1.

Table 1: Material properties and experimental parameters

| Young's Modulus | $E$ | 4.22 kPa |
|---|---|---|
| Poisson's Ratio | $\nu$ | 0.49 |
| Density | $\rho$ | 1,010 kg/m$^3$ |
| Stress-Optic Coefficient | $C$ | 3.3 x 10$^{-8}$ Pa$^{-1}$ |
| Wavelength of light | $\lambda$ | 540 nm |
| Sphere radius | $R$ | 7.5 mm |
| Applied force | $F$ | 9.8-98.1 mN in increments of 9.8 mN, 98.1-294.3 mN in increments of 19.6 mN |

## 2.2 Experimental setup

Figure 2 shows an image of the experimental setup [23]. The gelatin cuboid and acrylic container rest on top of a digital scale, which is used to measure the force applied to the top of the sphere. This is done to study the relationship between applied force and optical phase retardation. Under a planar load assumption, a purely elastic material will exhibit a linear relationship between applied force and phase retardation. For the Hertzian contact scenario, however, this is not the case, as a non-linear relationship between applied force and phase retardation is observed.

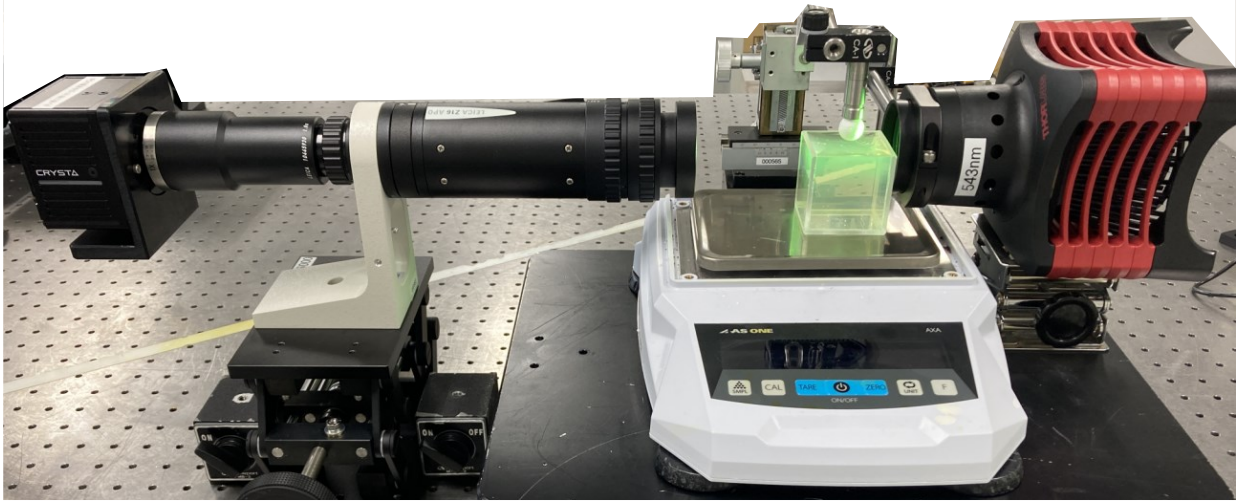

Figure 2: Image of experimental setup. A green light laser (540nm) is used to create coherent light, which first passes through a collimating lens, then through a linear polarizer (0°) and quarter waveplate (45°) to create left-handed circular polarized light. The light is incident upon the gelatin cuboid and emerges with an accumulated phase retardation $\Delta$ and orientation $\psi$ and is then recorded by a Photron Crysta PI-5WP high-speed polarization camera.

The gelatin cuboid is positioned in a circular polariscope with the following optical elements: a Thorlabs SOLIS-565C light source provides coherent monochromatic light of wavelength, $\lambda$ = 540 nm, which first passes through a plane polarizer whose transmission axis is horizontal and aligned with the x-axis shown in Fig. 1a. The light then passes through a quarter-wave plate positioned with its "fast" axis at +45º with respect to the x-axis. This creates left-handed, circular polarized light, which is then incident upon the gelatin cuboid. As the light propagates through the material, it accumulates phase retardations and rotations corresponding to the stress states encountered along the given light ray. It is assumed that the clear acrylic plates on either end of the gelatin do not alter either the phase or the orientation of the polarized light. Finally, the light emerges from the stressed model as elliptically polarized light and is recorded by a Photron Crysta polarization camera (Photron, CRYSTA PI-5WP), which features an array of "super-pixels." Each super-pixel is discretized into four quadrants, which measure? the polarization state at orientations of 0°, 45°, 90° and 135° with respect to the x-axis. The four polarizer orientations enable the linear Stokes parameters [23] to be identified, from which the degree and angle of linear polarization can be determined.

By applying the phase shifting method, where the analyzer is successively rotated through the orientations of $\theta = 0°$, 45°, 90° and 135°, four distinct output Stokes vectors are obtained [36, 37]. The output intensities of each of these Stokes vectors are denoted by $I_1$, $I_2$, $I_3$, and $I_4$, respectively, and are related by,

$$I_0 = \frac{I_1 + I_2 + I_3 + I_4}{2}. \tag{3}$$

After applying the phase shifting method to obtain the four Stokes vectors, and the corresponding light intensities, the phase retardation induced by the optically equivalent model is calculated as follows,

$$\Delta = \frac{\lambda}{2\pi} \sin^{-1} \frac{\sqrt{(I_3 - I_1)^2 + (I_2 - I_4)^2}}{I_0}, \tag{4}$$

while the output principal orientation of the light ellipse induced by the optically equivalent model is given by,

$$\psi = \frac{1}{2}\tan^{-1}\frac{(I_3 - I_1)}{(I_2 - I_4)}. \qquad (5)$$

Therefore, if the secondary principal stresses and orientations (see Section 3.2) are given, the phase retardation can be determined for the optically equivalent model by applying the phase shifting method. This provides a theoretical expectation for the results obtained in an integrated photoelasticity experiment. In order to obtain the theoretical phase retardation expected in an integrated photoelasticity experiment, the Hertzian stress field must be determined. The recorded light intensity fields are then post-processed using CRYSTA Stress Viewer (Photron Ltd.) to obtain phase retardation fields.

## 3. Hertzian contact theory

Hertzian contact theory has a history of about 140 years and it is extensively treated in both classical elasticity textbooks and specialized books and monographs, e.g. [6, 30, 38]. However, due to the complexity of the problem, typically only the solutions along the $z = 0$ plane and the $z$-axis (see Fig. 1a) are given in explicit form. In this work, the stress field everywhere in the substrate is needed. Hence a review of the theory is included, for completeness.

To determine the stress state in a material subject to Hertzian contact, where a rigid sphere of diameter $D$ is pressed into the top of a flat surface with force $F$, a theoretical model is derived [6, 30, 38, 39]. In this model, the sphere is rigid while the substrate material bears all deformation. In the present experiment, the sphere is made of styrene, which has an elastic modulus of approximately 2 GPa >> 4.22 kPa for gelatin. The theoretical analysis is greatly simplified if axial symmetry is assumed, along with allowing the substrate material to extend to a semi-infinite domain, see Fig. 1b. This analysis also assumes that the material is linearly elastic and that the small strain approximation is valid, despite the large displacements expected with such a soft material like gelatin. The contact between the sphere and substrate is assumed to be frictionless; therefore, only a normal pressure is transmitted by the sphere to the substrate.

## 3.1 Derivation of stress and displacement fields

Consider an elastic half-space in cylindrical coordinates, which extends from $0 \leq z < \infty$, and $0 \leq r < \infty$, of Young's Modulus, $E$ = 4.22 kPa and Poisson's Ratio, $\nu$ = 0.49. The half-space is subject to a Hertzian pressure distribution on the $z = 0$ plane, along $0 \leq r \leq a$, where $r = a$ is the contact radius, see Fig. 1b. The normal pressure distribution corresponding to a sphere loaded along the epicentral $z$-axis, with force $F$, and radius $R = D/2$, is given by,

$$\sigma_{zz}(r, z = 0) = p_0\sqrt{a^2 - r^2}, \quad \text{for } 0 \leq r \leq a, \tag{6a}$$

$$\sigma_{zz}(r, z = 0) = 0, \quad \text{for } r > a, \tag{6b}$$

where,

$$p_0 = \frac{3F}{2\pi a^2}, \tag{7}$$

is the maximum pressure [6]. The contact radius, $a$ is given by,

$$a = \left(\frac{3FR}{4E^*}\right)^{1/3}, \qquad (8)$$

where $E^*$ is the Effective Modulus, see Eq. (2). Since frictionless contact is assumed, $\sigma_{rr} = \sigma_{\theta\theta} = \sigma_{rz} = 0$ on $z = 0$. The sphere contacts the surface of the half-space with a constant force, $F$; therefore, static equilibrium equations are employed and are written as follows,

$$\frac{\partial \sigma_{rr}}{\partial r} + \frac{\sigma_{rr} - \sigma_{\theta\theta}}{r} + \frac{\partial \sigma_{rz}}{\partial z} = 0, \qquad (9)$$

$$\frac{\partial \sigma_{rz}}{\partial r} + \frac{\sigma_{rz}}{r} + \frac{\partial \sigma_{zz}}{\partial z} = 0. \qquad (10)$$

Notice that body forces are absent from Eqs. (9) and (10). Gravity induces a hydrostatic stress to the material, which does not alter the secondary principal stress difference. Therefore, gravity does not affect the phase retardation in the photoelastic measurements and will be ignored in this work.

In order to ensure a physically meaningful displacement field, the stress compatibility relations are utilized (in cylindrical coordinates) and are given by,

$$\nabla^2 \sigma_{rr} - \frac{2}{r^2}(\sigma_{rr} - \sigma_{\theta\theta}) + \frac{1}{1+\nu}\frac{\partial^2 e}{\partial r^2} = 0, \qquad (11)$$

$$\nabla^2 \sigma_{\theta\theta} - \frac{2}{r^2}(\sigma_{rr} - \sigma_{\theta\theta}) + \frac{1}{1+\nu}\frac{1}{r}\frac{\partial e}{\partial r} = 0, \tag{12}$$

$$\nabla^2 \sigma_{zz} + \frac{1}{1+\nu}\frac{\partial^2 e}{\partial r^2} = 0, \tag{13}$$

$$\nabla^2 \sigma_{rz} - \frac{\sigma_{rz}}{r^2} + \frac{1}{1+\nu}\frac{\partial^2 e}{\partial r \partial z} = 0, \tag{14}$$

where $e = tr(\bar{\sigma}) = \sigma_{rr} + \sigma_{\theta\theta} + \sigma_{zz}$. Love (1929) derived a biharmonic stress function $\xi = \xi(r,z)$, which identically satisfies the equilibrium equations, as well as the compatibility relations [39]. The stress components (in cylindrical coordinates) can be derived from the Love stress function as follows,

$$\sigma_{rr} = \frac{\partial}{\partial z}\left[\nu \nabla^2 \xi - \frac{\partial^2 \xi}{\partial r^2}\right], \tag{15}$$

$$\sigma_{\theta\theta} = \frac{\partial}{\partial z}\left[\nu \nabla^2 \xi - \frac{1}{r}\frac{\partial^2 \xi}{\partial r}\right], \tag{16}$$

$$\sigma_{zz} = \frac{\partial}{\partial z}\left[(2-\nu)\nabla^2 \xi - \frac{\partial^2 \xi}{\partial z^2}\right], \tag{17}$$

$$\sigma_{rz} = \frac{\partial}{\partial r}\left[(1-\nu)\nabla^2 \xi - \frac{\partial^2 \xi}{\partial z^2}\right]. \tag{18}$$

Similarly, the radial and axial displacements in the half-space can be determined using the Love stress function and are given by,

$$u_r = -\frac{1+v}{E}\frac{\partial^2 \xi}{\partial r \partial z}, \tag{19}$$

$$u_z = \frac{1+v}{E}\left[2(1-v)\nabla^2 \xi - \frac{\partial^2 \xi}{\partial z^2}\right], \tag{20}$$

respectively. To determine the Love stress function the biharmonic equation is solved, which is given by,

$$\nabla^4 \xi = \nabla^2 \nabla^2 \xi = 0. \tag{21}$$

The Hankel transform method [40] is used to solve this equation while enforcing the boundary conditions given by Eq. (6). The resulting Love stress function for the Hertzian contact problem is then,

$$\xi(r,z) = \int_0^\infty f(k)\left[\left(\frac{2v}{k}+z\right)\frac{J_0(kr)}{k^2}\right]e^{-kz}dk, \tag{22}$$

with

$$f(k) = p_0 \frac{(\sin ak - ak \cos ak)}{ak^2}, \tag{23}$$

where $J_0$ specifies the Bessel function of first kind of order zero, and $p_0$ and $a$ are given by Eqs. (7) and (8), respectively. The dummy variable of integration, "$k$", is termed the wavenumber. The Love stress function is not provided as a closed form solution; despite this, the function can be approximated numerically by using a sufficiently large number for $k$. Using the Love stress function along with Eqs. (15-18), the stress components everywhere in the domain are given explicitly as follows,

$$\sigma_{rr}(r,z) = \int_0^\infty f(k)\left[(1-kz)J_0(kr) + (2v-1+kz)\frac{J_1(kr)}{kr}\right]e^{-kz}dk, \quad (24)$$

$$\sigma_{\theta\theta}(r,z) = \int_0^\infty f(k)\left[2vJ_0(kr) + (1-2v-kz)\frac{J_1(kr)}{kr}\right]e^{-kz}dk, \quad (25)$$

$$\sigma_{zz}(r,z) = \int_0^\infty f(k)[(1+kz)J_0(kr)]e^{-kz}dk, \quad (26)$$

$$\sigma_{rz}(r,z) = \int_0^\infty f(k)[2(1-v)J_1(kr) + (2v-2+kz)J_1(kr)]e^{-kz}dk. \quad (27)$$

where $J_1$ specifies the Bessel function of first kind of order one.

The Love stress function is also used to determine the displacements with Eqs. (19) and (20), and are given as,

$$u_r(r,z) = -\frac{v+1}{E}\int_0^\infty f(k)\left[(2v-1+kz)\frac{J_1(kr)}{k}\right]e^{-kz}dk, \quad (28)$$

$$u_z(r,z) = \frac{v+1}{E} \int_0^\infty f(k)\left[(2v - 2 - kz)\frac{J_0(kr)}{k}\right] e^{-kz} dk .\qquad(29)$$

With the stress state inside the half-space fully defined, the secondary principal stresses on planes perpendicular to the light propagation direction, i.e., the $y$-axis, see Fig. 1, can now be determined.

## 3.2 Secondary principal stress

The secondary principal stresses are the max. and min. normal stresses in the plane orthogonal to the light propagation direction for any in-plane orientation of the stress element at a point, and are used in calculating the optically equivalent model [23]. In the schematic shown in Fig. 1a, the light propagates along straight lines parallel to the $y$-axis, so that the secondary principal stresses have to be computed on planes parallel to the $x$-$z$ one. It is, therefore, more convenient to determine the secondary principal stresses from a Cartesian coordinate system, by suitable rotations of the stress tensor. Then,

$$\sigma_{1,2} = \frac{\sigma_{xx} + \sigma_{zz}}{2} \pm \sqrt{\left(\frac{\sigma_{xx} - \sigma_{zz}}{2}\right)^2 + \sigma_{xz}^2} .\qquad(30)$$

The orientation of secondary principal stresses is given by,

$$\tan(2\phi) = \frac{\sigma_{xz}}{\sigma_{xx} - \sigma_{zz}}, \tag{31}$$

where $\phi$ is measured with respect to the $x$-axis, see Fig. 1c. Note that in this analysis, only the in-plane (i.e., $x$-$z$ plane, see Fig. 1a) stresses are considered to contribute to the retardation (i.e., to stress-induced optical anisotropy). No effect of stresses with components along the $y$-axis is considered. This is in line with the understanding that birefringence is induced by the stressing of a thin lamina in its plane and not by out-of-plane stresses. What this assumption ignores is that the out-of-plane stressing contributes to the thinning or thickening of the lamina, which may contribute to the retardation in this way.

## 3.3 Boundary effects

The axisymmetric elastic half-space is segmented into a cuboid with dimensions equal to the dimensions of the experimental cuboid in Fig. 1a. This segmentation exposes tractions on each of the side walls as well as the bottom surface. In the experiment, it is likely that friction exists between the gelatin and acrylic container during the Hertzian loading. This boundary condition will influence the stress state in the cuboid. To ensure that these tractions and boundary conditions are small in comparison to the Hertzian contact stress, the force applied to the sphere must be limited. Reasonable approximations can be obtained if the boundary is at least four contact radii away from the center of loading, as it is confirmed later in Section 5.2.

After the half-space is segmented into a cuboid, the secondary principal stresses and orientations are used in conjunction with Eq. (4) to determine the phase retardation in the optically equivalent model. This provides a theoretical basis to compare to the experiments. Furthermore, the phase retardation can also be calculated using the optically equivalent model with stress states obtained through numerical simulations. The advantage of the latter is that they can relax the small strain and small contact patch assumptions inherent to classical Hertz theory, but perhaps inapplicable to the soft material in hand. In this way, the limitations of using the classical Hertzian contact theory on soft materials can be assessed.

## 4. Numerical simulation of Hertzian contact

The goal of the simulations is to obtain the state of stress inside a material subject to Hertzian contact and to compare the simulated phase retardation field to theory and experiments. Furthermore, the simulation considers finite strains, which is analytically intractable but relevant to the soft material of this study, and assess the extent that the infinitesimal strain assumption negatively affects the results.

An axisymmetric domain is used for the simulations, where a rigid sphere is pressed into the top surface, see Fig. 3. The simulations are performed with the commercial non-linear code Abaqus/Standard v. 6.14 (implicit), using a fine, rectangular, 0.1 x 0.1 mm mesh, see Fig. 3, with 4-node, reduced-integration, first-order, axisymmetric solid elements (CAX4R). This mesh density was arrived at after suitable parametric studies. Linearly elastic material behavior is assumed, with the same material properties used throughout this work, see Table 1. The contact between the sphere and material is frictionless. A simulation was performed with a friction coefficient of 0.1, however, these results were nearly identical to the frictionless results. The axial and radial length of the axisymmetric domain is 60 x 60 mm$^2$, which is slightly larger than the experimental domain so that boundary effects do not affect the results. The simulations are performed under force control. See Table 1 for the problem parameters and range of applied forces.

Figure 3 shows the simulation results on the $r$-$z$ plane in terms of the von Mises equivalent stress, including a zoomed-in view of the contact region where the largest von Mises stress occurs. This is below the surface and is 959 Pa for a sphere pressed with $F$ = 49 mN. It is apparent from Fig. 3 that most of the stress is localized in the subsurface region directly below the sphere, as is well known ([6], [30], [38]). Indeed, more than 2 diameters away the stress field has decayed to almost zero.

The secondary principal stresses and orientations are obtained as before, using Eqs. (30) and (31). This provides the stresses needed to determine the phase retardation in the optically equivalent model [23], which can then be compared to its theoretical and experimental counterparts. For this calculation, the axisymmetric simulation domain, i.e., cylinder, is segmented into a cuboid of dimensions 44 x 44 x 47 mm$^3$, as in the schematic shown in Fig. 1a.

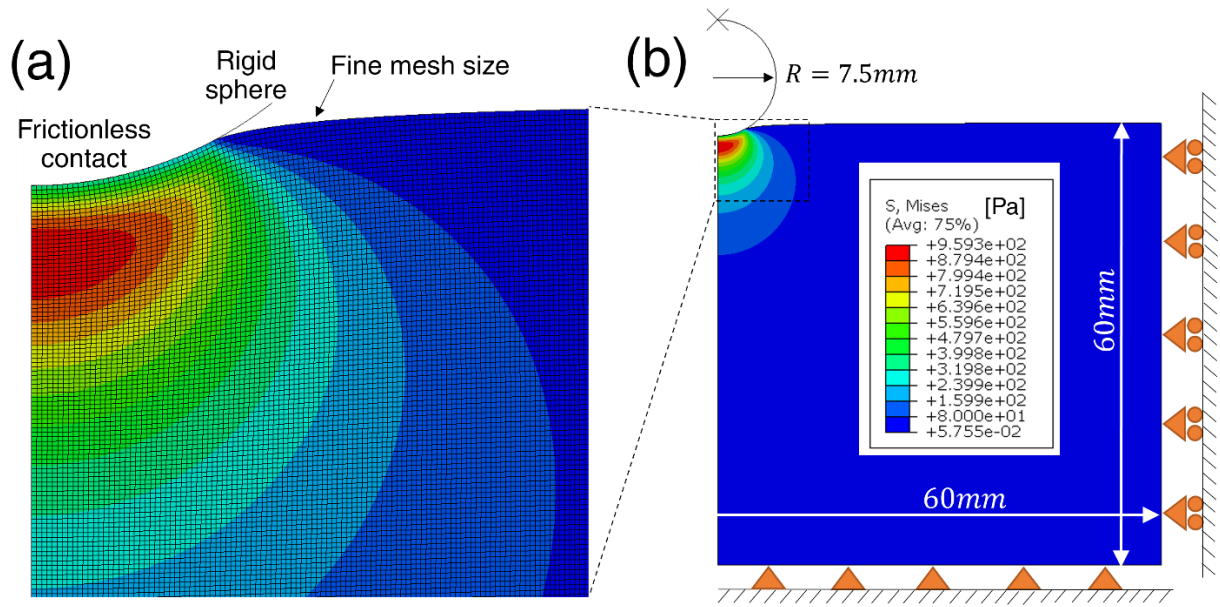

Figure 3: (a) Zoomed-in view of loading region depicting the Hertzian contact and FEA mesh-size. (b) Axisymmetric Abaqus model showing equivalent stress distribution on the r-z plane for F = 49 mN case.

## 5. Results and discussion

This section examines the phase retardation fields of the experiment, theory, and numerical simulations, along with the maximum equivalent stress experienced by the gelatin. The maximum axial displacements and contact patch radii of the Hertzian loading are also explored.

### 5.1 Phase retardation fields

Integrated photoelasticity for soft material was carefully evaluated in [23]. Utilizing the method we developed in that work, we can obtain phase retardation fields from the stress fields, which are analytically and numerically calculated, in a soft material.

Using the experimental setup described in Section 2.2, phase retardation fields are acquired for the applied forces listed in Table 1. Each of these forces is used to calculate the theoretical and numerical stress states, which are then used to determine the optically equivalent model [23] and, thus, the theoretical and numerical phase retardation fields. Figures 4-7 show the phase retardation fields for the $F$ = 9.8, 49.1, 98.1, and 157.0 mN loading scenarios, respectively. Each figure contains an (a) experimental, (b) theoretical, and (c) numerical phase retardation field. The interrogation windows range from $0 \leq z \leq 10$ mm, and $-10 \leq x \leq 10$ mm, (normalized window of $0 \leq z/R \leq 1.33$, and $-1.33 \leq x/R \leq 1.33$), which is outlined in Fig. 1a as a dashed black line. This is also the domain used to determine the theoretical and numerical phase retardation fields. The origin of this domain is the point of first contact between the sphere and the gelatin. The viewing axis of the camera lens is vertically aligned and parallel with the top surface of the gelatin, i.e., $z$ = 0. The sphere is progressively submerged below this plane as the force increases in Figs. 4-7.

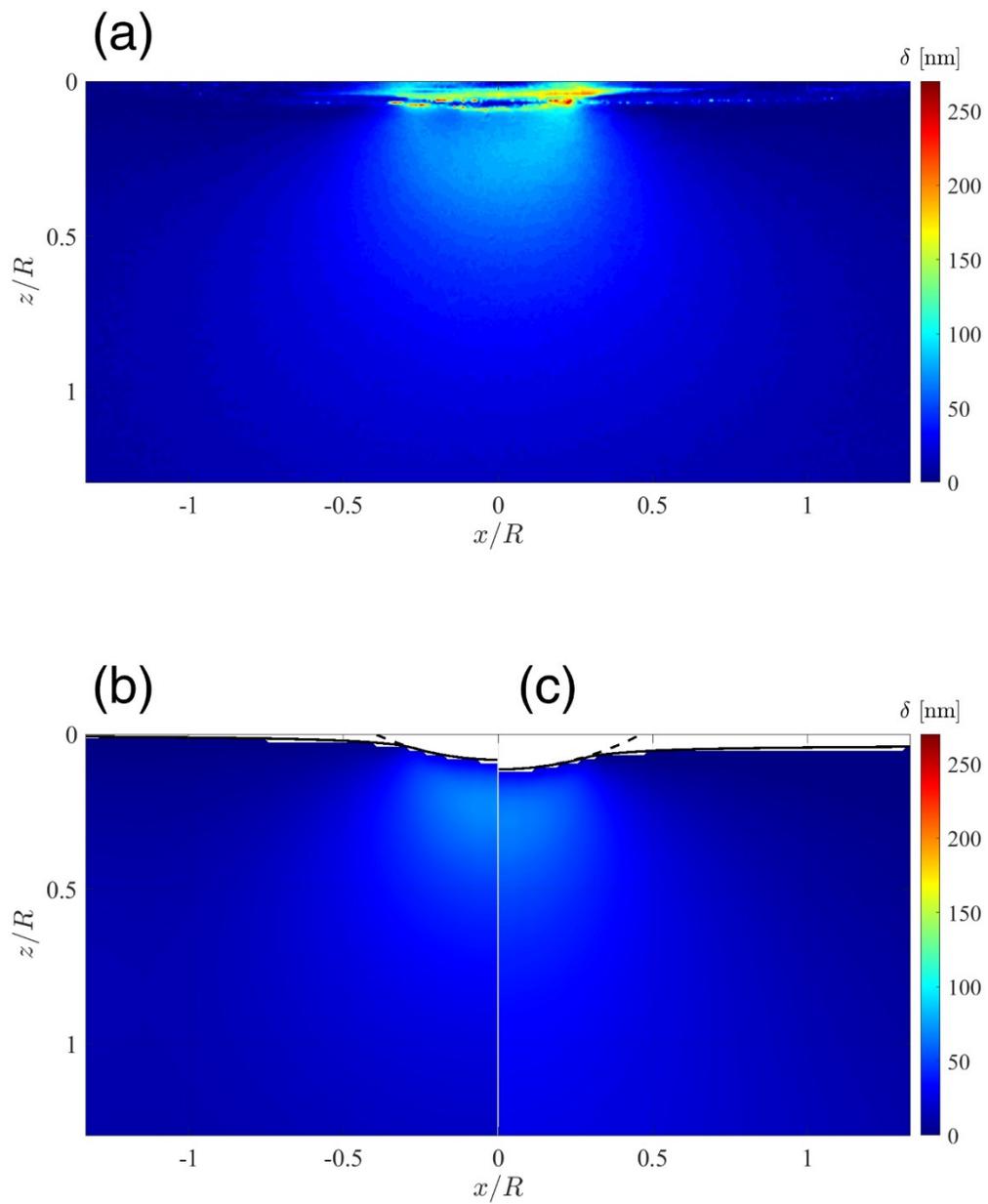

Figure 4: (a) Experimental, (b) theoretical, and (c) numerical phase retardation field for the $F$ = 9.8 mN loading case.

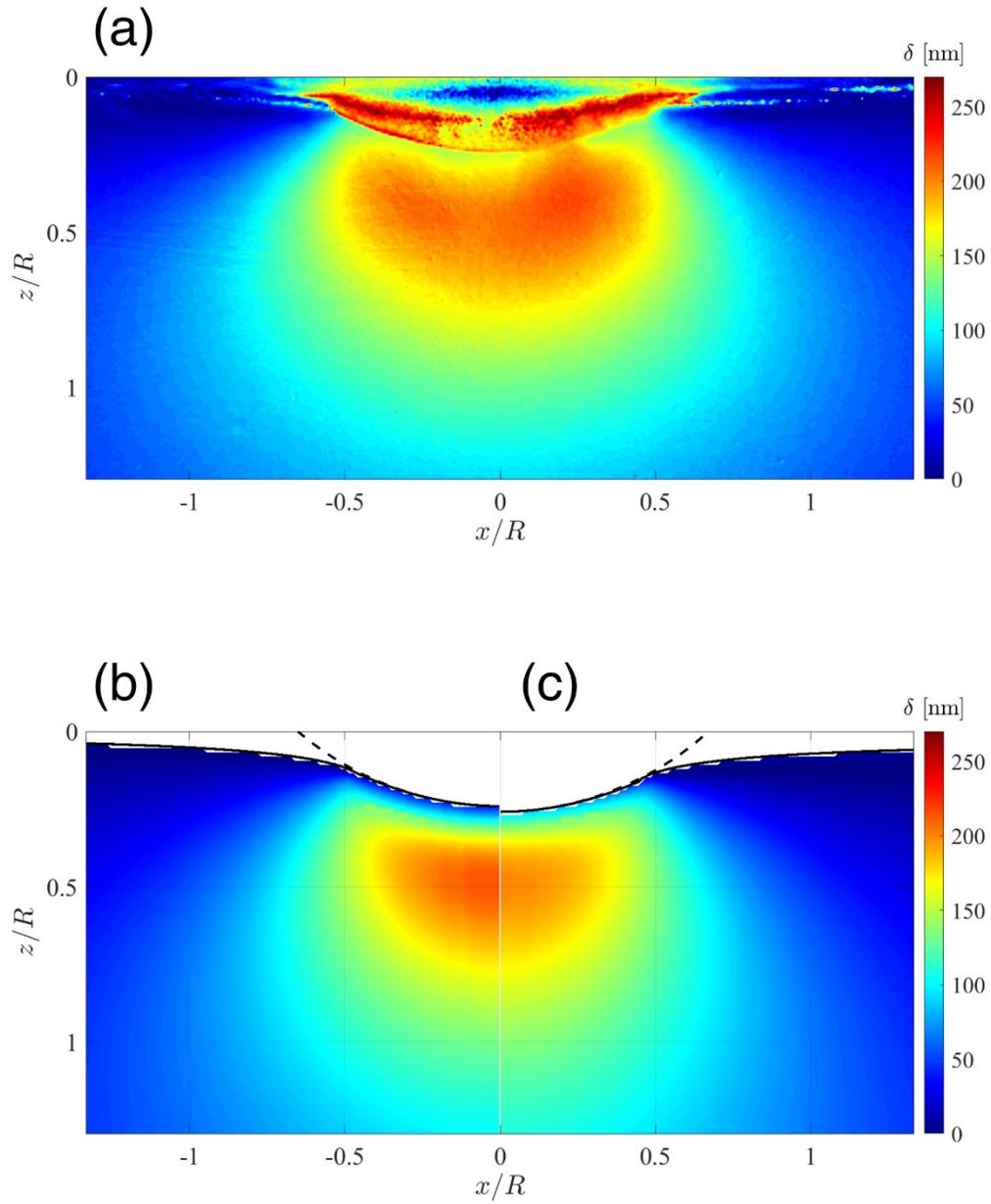

Figure 5: (a) Experimental, (b) theoretical, and (c) numerical phase retardation field for the $F$ = 49.1 mN loading case.

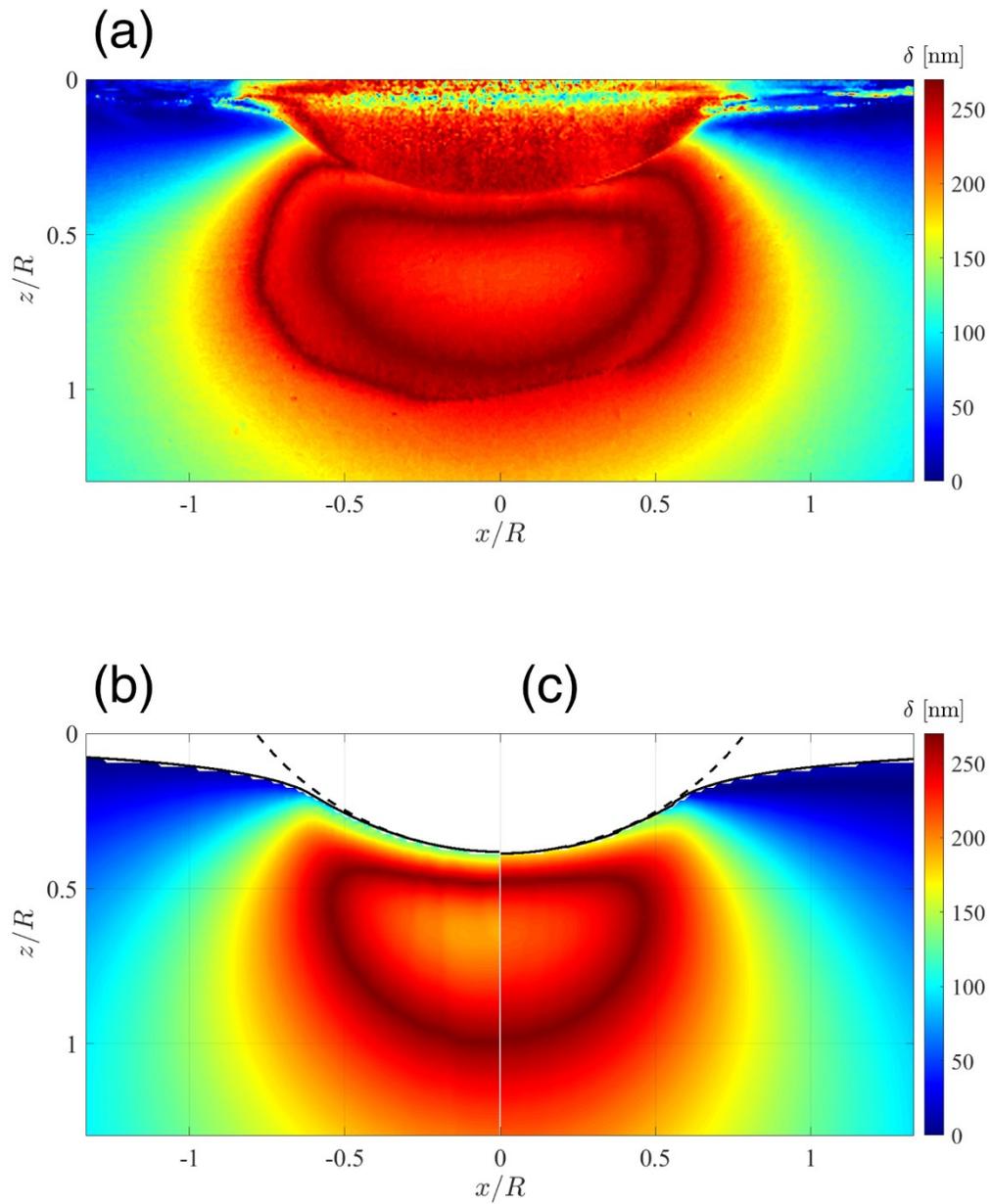

Figure 6: (a) Experimental, (b) theoretical, and (c) numerical phase retardation field for the $F$ = 98.1 mN loading case.

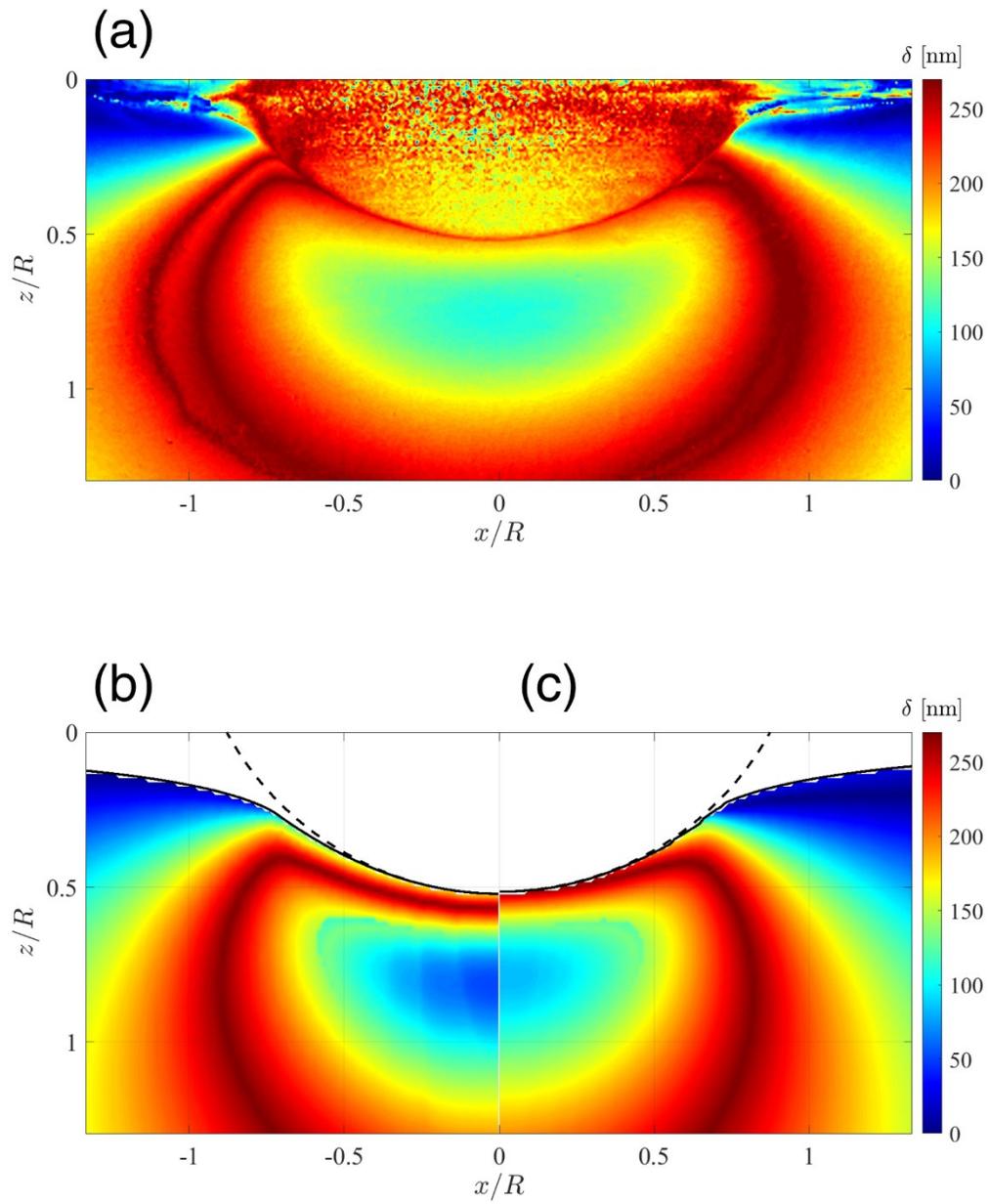

Figure 7: (a) Experimental, (b) theoretical, and (c) numerical phase retardation field for the F = 157.0 mN case.

All of the figures reveal an approximately semicircular-shaped region of phase retardation in the center, underneath the sphere. The dark blue and dark red extremes correspond to the bounds of the experimental phase retardation field, which are 0 and 270 nm, respectively. The upper limit corresponds to one half of the wavelength of the light used, i.e., $\lambda/2$ = 270 nm. This range of phase retardation, i.e., $0 \leq \delta < 270$ nm, corresponds to an angular phase retardation of $0 \leq \Delta < \pi$. The experiments are limited to this range since, in photoelasticity, the emergent light is interpreted as major and minor electric field strengths on a light ellipse that range from 0 to $\pi$. When the actual phase retardation induced by the stressed material exceeds 270 nm, i.e., $\pi$, the measured phase will be recorded at a value less than 270 nm. This phenomenon is known as phase wrapping and can cause ambiguity in interpretation of measurement results. For the loading scenarios represented in Figs. 4 and 5, the maximum phase does not exceed 270 nm; therefore, phase wrapping is not present. However, in Figs. 6 and 7 phase wrapping exists and must be taken into account. In Fig. 6a, it can be seen that the maximum phase reaches 270 nm, i.e., red colored ring, while a relatively smaller phase retardation, indicated by the light red color, exists in the region inside the ring, below the sphere. It is here where care must be taken in analyzing the results. The light red region under the sphere, and inside the dark red ring, is interpreted as the highest phase retardation, which, according to the scale, is approximately 220 nm. For this scenario, the unwrapped maximum phase retardation is calculated as $\delta^{max}$ = 270 + (270-220) = 320 nm. A similar phase wrapping situation exists for the theoretical and numerical fields of Figs. 6b and c. Here, the unwrapped maximum phase retardations are 348 nm and 326 nm, respectively. This identifies fair agreement between the maximum phase retardation of the experiment, and its theoretical and numerical counterparts. The semicircular shaped contours and their relative positions in Figs. 4-6 also show good qualitative agreement. This suggests that the theoretical and numerical results are capturing the phase retardation induced by the Hertzian contact observed in the experiments.

The phase retardation fields of the $F$ = 157.0 N force case, shown in Fig. 7, begin to identify discrepancies between experiments, theory, and numerics. Although they share the same general shape and qualitative characteristics, deviations in maximum phase retardation become apparent inside the dark red rings. It is possible that, in the experiments, the stress field is extending toward the container boundaries resulting in an alteration of the Hertzian contact. Furthermore, the contact radius is becoming large and close to the radius of the sphere, which violates the Hertzian contact assumptions. These results, shown in Fig. 7, are characteristic of large forces and displacements.

The maximum (unwrapped) phase retardation is determined for each of the fields shown in Figs. 4-7, as well as for the theoretical and experimental tests listed in Table 1. Figure 8 shows the relationship between maximum phase retardation and applied force for the experiment (red circles), theory (dashed blue line), and numerical simulations (green triangles). It is evident that the relationship between maximum phase and applied force is non-linear. This relationship is examined, in detail, in the following section.

For small force cases, i.e., $F <$ 100 mN, excellent agreement is observed in Fig. 8 between experiment, theory, and numerical simulations. This indicates that classical Hertzian theory and/or numerical models can be used for interpreting the stress state inside the gelatin, when the contact patch is much smaller than the sphere radius.

However, for higher force cases, the experimental phase retardation deviates from the theoretical and numerical phase retardations. Again, this deviation is likely due to forces being high enough to make the contact radius approach that of the sphere, inducing large deformations and rendering the Hertzian contact approximations invalid.

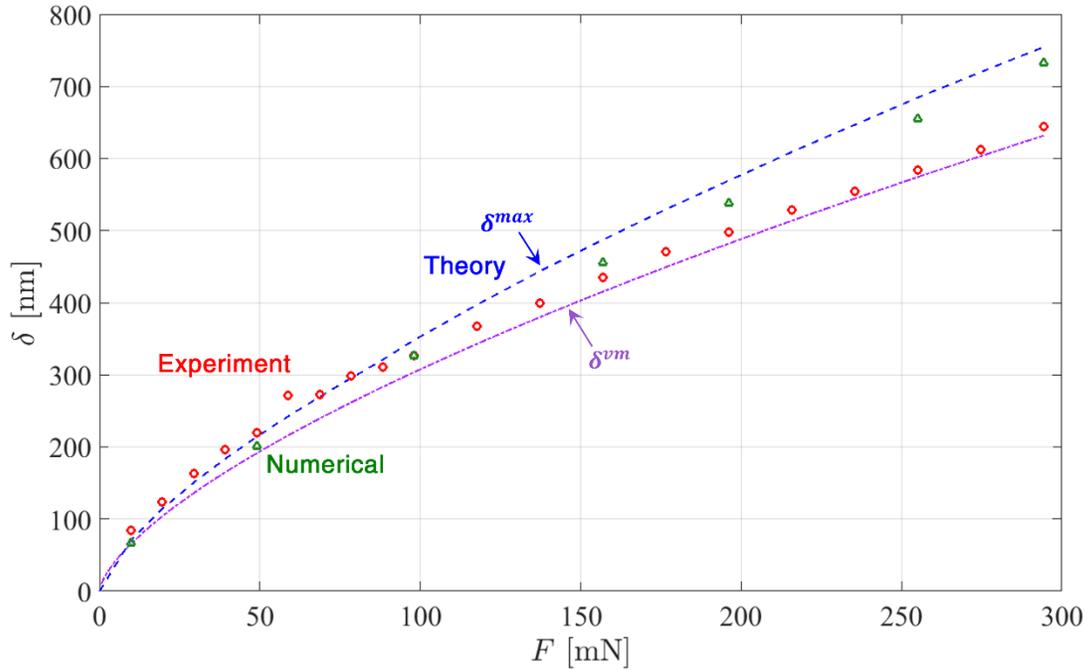

Figure 8: Relationship between maximum phase retardation and applied force for experiment (red circles), theory (blue dashed line), and numerical (green triangles). The phase retardation passing through the point of maximum von Mises stress, given by Eq. (34), is presented with a dash-dotted purple line.

As stated earlier, the inverse problem of integrated photoelasticity is ill-posed. This implies that stress fields can be determined in a tomographic sense solely from an experimental phase retardation field for the entire domain only in special cases. Phase retardation fields provide insight into the mechanics and identify areas of maximum phase retardation; however, the most important parameter, with respect to the onset of yielding, is the equivalent stress, which we discuss in the next subsection.

## 5.2 Principal stress components and maximum equivalent stress

In each of the loading scenarios, the maximum phase retardation occurs below the surface and along the $z$-axis. Due to the axisymmetry, $\sigma_{zz} = \sigma_1$ and $\sigma_{xx} = \sigma_{yy} = \sigma_2$ everywhere along the $z$-axis [6], the von Mises equivalent stress is reduced to,

$$\sigma_{VM} = |\sigma_1 - \sigma_2|. \tag{32}$$

This relationship holds everywhere along the $z$-axis, which includes the point of maximum von Mises stress, which occurs at approximately $z = 1.55a$. For light rays passing through the $z$-axis, i.e., traveling on the $x = 0$ plane (see Fig. 1), the Integral Wertheim Law [23] is applicable, as there is no rotation of secondary principal directions along this plane. Also, since along the $x = 0$ plane there is no $\sigma_{xz}$ shear, the secondary principal stresses reduce to $\sigma_1 = \sigma_{zz}$, and $\sigma_2 = \sigma_{xx} = \sigma_{\theta\theta}$. With these variables it may seem like the Wertheim integral is reduced in complexity; however, a closed form solution of the integral does not exist, to the authors' knowledge.

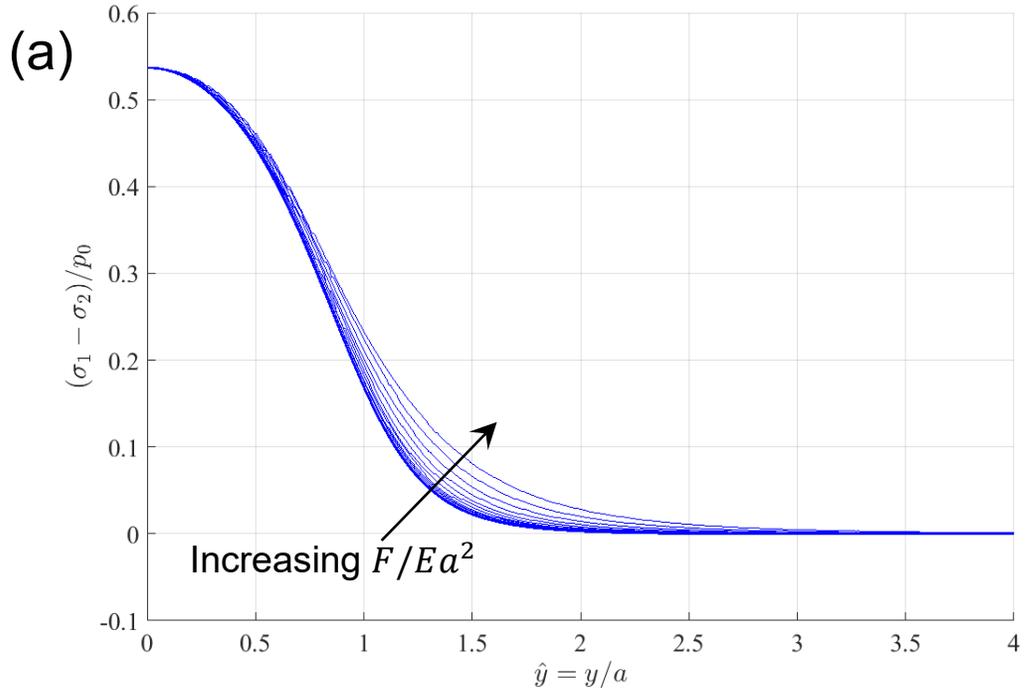

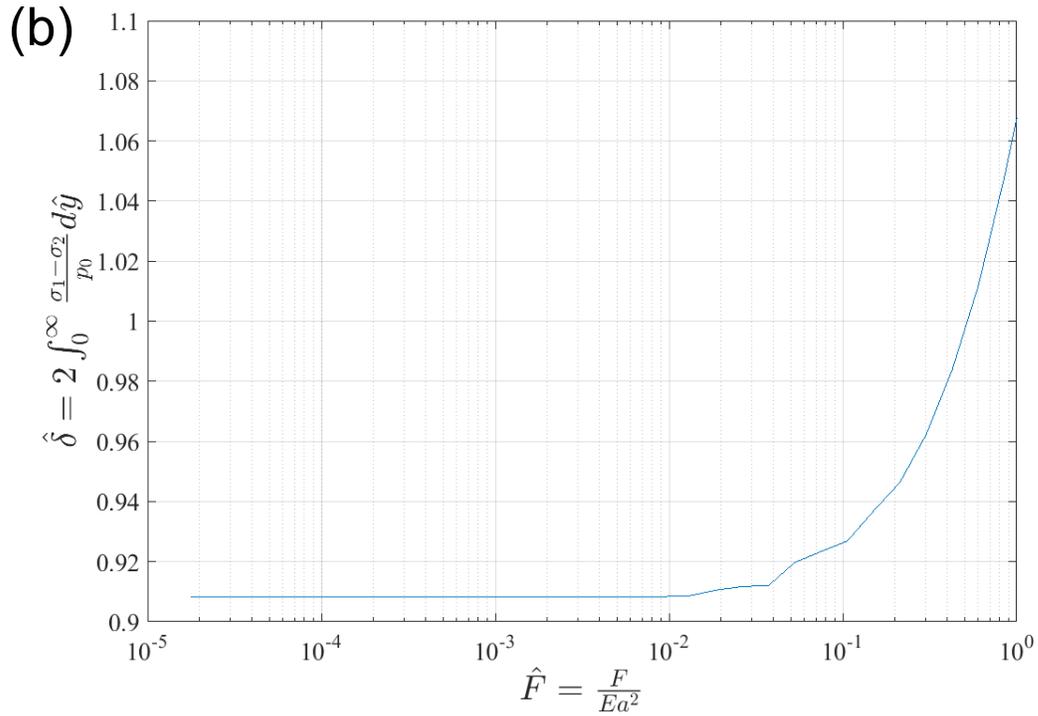

Figure 9: (a) Secondary principal stress profile along the light propagation direction, $\hat{y} = y/a$, of rays passing through the point of maximum von Mises stress $\sigma_{vm}$, for a range of forces, i.e., $10^{-5} < F/Ea^2 < 1$. For small forces, i.e., $F/Ea^2 \ll 1$, stress profiles are self-similar. (b) Normalized area under the stress profiles shown in (a), where a constant value, i.e., $\hat{\delta} = 0.91$, is observed for $F/Ea^2 \ll 1$.

To examine the behavior of this integral, stress profiles are plotted for a variety of loading cases, $10^{-5} < F/Ea^2 < 1$, in Fig. 9a, for light rays passing through the point of maximum von Mises stress. For small forces, i.e., $F/Ea^2 \ll 1$, the stress profiles are invariant, revealing self-similarity. However, for increasing forces, $F/Ea^2 \to 1$, the stress profiles deviate from the self-similar profile, and show a greater secondary principal stress difference in the range $1 < \hat{y} < 2$, see Fig. 9a. The normalized area under each of these profiles, which is given by,

$$\hat{\delta} = \frac{\delta}{c p_0 a} = 2 \int_0^\infty \frac{(\sigma_1 - \sigma_2)}{p_0} d\hat{y}, \tag{33}$$

is shown in Fig. 9b for each load case. It is apparent that for small forces, i.e., $F/Ea^2 \ll 1$, the integral value is a constant $\hat{\delta} = 0.91$, while for large ones, $F/Ea^2 \to 1$, the integral value is ever-increasing. For these large forces, the contact radius approaches the sphere radius, where the Hertzian contact approximation becomes invalid. For the small forces, however, the constant $\hat{\delta} = 0.91$ identifies a relatively simple relationship between the phase retardation passing through $\sigma_{vm}^{max}$ and applied force, which is given by,

$$\delta^{vm} = 0.91 \frac{c}{\pi} \left( \frac{9 E^* F^2}{2R} \right)^{1/3}. \tag{34}$$

This equation is plotted in Fig. 8 as a dash-dotted purple line. It is apparent that this line qualitatively follows the data trends, but deviates from Hertz's theory (blue dashed line) with increasing force. In that case, the stress profiles, passing through $\sigma_{vm}^{max}$, deviate from the self-similar profile, producing a larger integral value, i.e., Eq. (33), and thus, yield a phase retardation greater than that predicted by Eq. (34). Hence, Eq. (34) is only applicable in the small force limit, i.e., $F/Ea^2 \to 0$. Nonetheless, Eq. (34) provides insight into the relevant parameters of the Hertzian, integrated photoelasticity problem. For example, it is evident that phase retardation scales with force to the 2/3$^{rd}$ power. Using this relation, one could use Hertzian loading in integrated photoelasticity as a load cell (within the appropriate force approximation).

Using Eq. (32), the maximum von Mises stress is calculated for each loading case, using the theoretical and numerical stresses. The axial position of maximum von Mises stress is approximately the same axial position as the maximum phase retardation. Figure 10 shows the relationship between maximum von Mises stress (which is also the maximum secondary principal stress difference) and maximum phase retardation for each of the loading conditions. The theoretical data points are presented with blue dashed line, while the numerical data points are shown with green triangles. For increasing force, the numerical results deviate from theory. This may be attributed to the difference in loading conditions between theory and numerics, as the theory uses a pressure boundary condition while the numerical simulation uses rigid contact, i.e., displacement boundary condition.

One may be tempted to use the experimental results and the classic "2D" stress optic law to obtain the principal stress difference; however, this would yield erroneous results for a 3D case, as they would produce the averaged principal stress difference across the depth of the cube. Another method is needed to make use of the present experimental results.

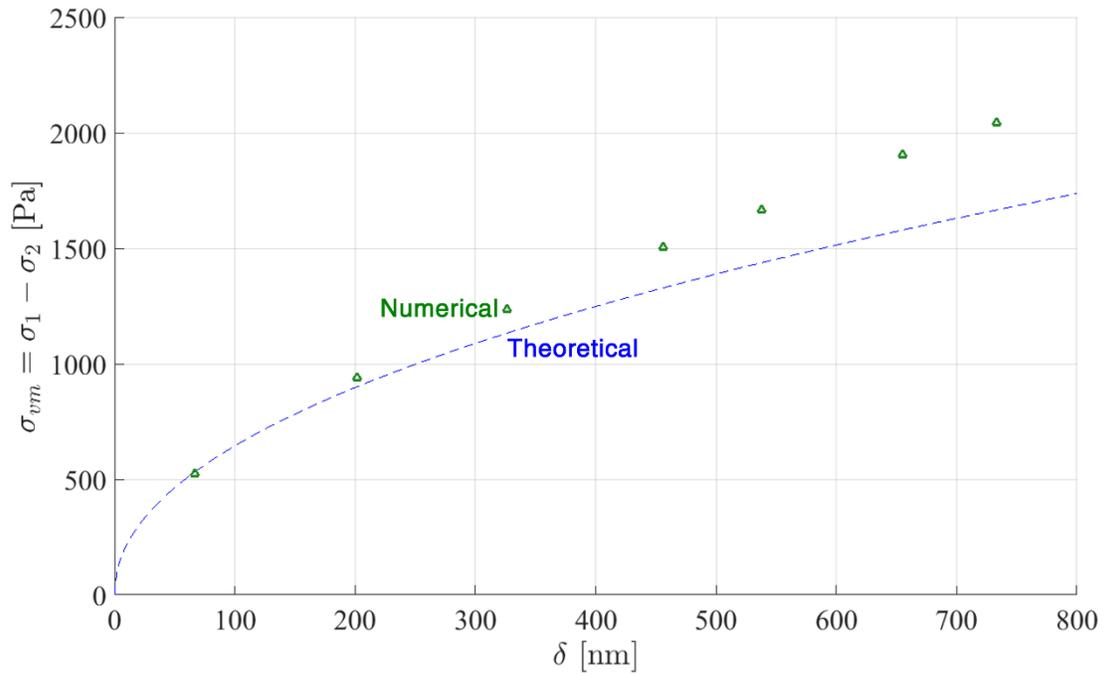

Figure 10: Relationship between maximum von Mises stress and the maximum phase retardation, evaluated on the y = 0 mid-plane, according to theory (blue dashed line) and numerical simulation (green). A fit line is applied to the theoretical data, given by Eq. (35).

In order to use the experimental data for stress reconstruction, (specifically, using the maximum phase retardation to calculate the maximum von Mises stress), the theoretical fit line in Fig. 10 can be employed which is given by,

$$\sigma_{vm} = (\sigma_1 - \sigma_2) = a\delta^n,  \qquad (35)$$

where fit parameters $a$ and $n$, are 71.9 Pa/nm and 0.48, respectively. Therefore, if given only a maximum phase difference, the maximum von Mises stress can be determined using this relationship. Hence, information about the stress state can be obtained directly from the experiments.

The von Mises stress is an important parameter when identifying the maximum stress state; however, it is also advantageous to know each individual principal stress component. Equation (35) cannot provide each principal stress component directly (only their difference is provided). Another equation is needed to determine each stress value. Here, it is recognized that the ratio between first and second principal stress components can be evaluated using Eqs. (24) and (26) at any point along the z-axis. This relationship is evaluated at the axial position where von Mises is maximum, and is given by,

$$\frac{\sigma_1}{\sigma_2} = \frac{\int_0^\infty f(k)[1 + kz]e^{-kz}dk}{\int_0^\infty f(k)\left[\frac{1}{2}(1 - kz) + v\right]e^{-kz}dk}. \tag{36}$$

This ratio is invariant with respect to $E$, although is dependent on $v$. However, since gelatin exhibits nearly incompressible behavior, the Poisson ratio is almost always $v = 0.49$. In the limit of small forces, i.e., $F/Ea^2 \ll 1$, Eq. (36) indicates that this ratio is constant, $\sigma_1/\sigma_2 = 3.3$. By using this ratio, in conjunction with Eq. (32), the first and second principal stresses can be determined, and since $\sigma_3 = \sigma_2$ everywhere along the z-axis, all three principal stresses are known at this location. This fully defines the stress state where the von Mises stress is maximum, providing crucial insight for yield forecasting.

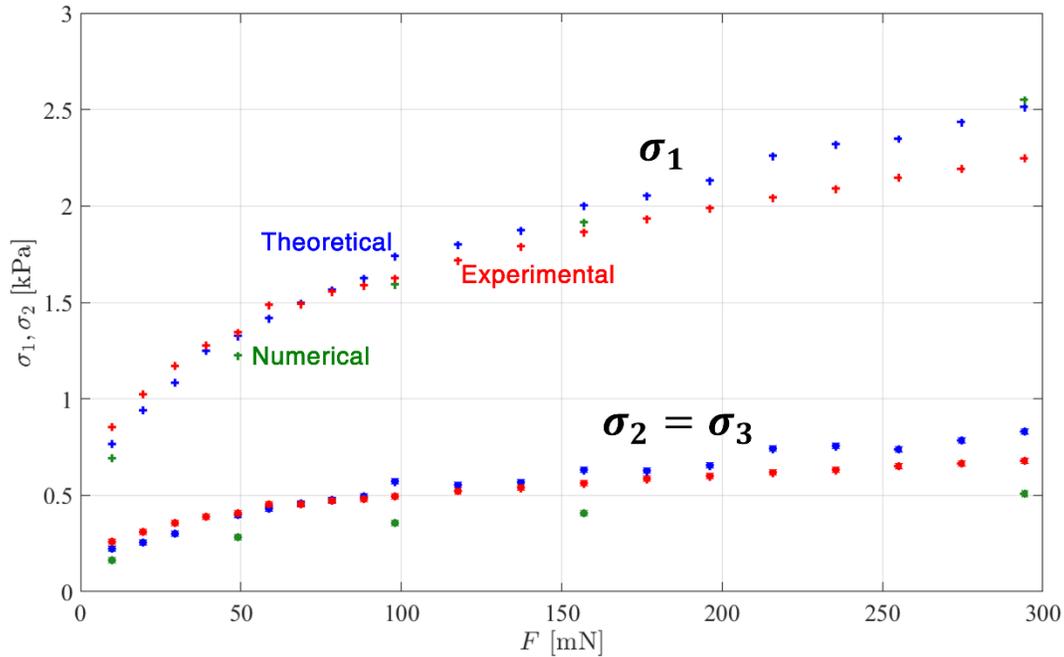

Figure 11: Principal stress components at the point of maximum von Mises stress for theory (blue), experiments (red) and numerical simulation (green). 'Plus' symbols are used for $\sigma_1$, while 'dots' are used for $\sigma_2$, respectively.

Using the experimental maximum phase retardation $\delta^{max}$, the von Mises stress is calculated using Eq. (35), and using the stress ratio of 3.3, the first and second principal stress components are determined for each of the loading cases, which are plotted in Fig. 11. Blue, red and green crosses represent the theoretical, experimental and numerical first principal stresses, respectively, while blue, red and green dots represent the theoretical, experimental and numerical second principal stresses, respectively. Satisfactory agreement is observed between theory and experiment for each stress component. This demonstrates that, by determining the phase retardation through experimentation, one can fully determine the stress state at the most critical point within the material.

## 5.3 Maximum surface displacement and contact radius

Another important parameter in the Hertzian contact problem is the displacement of the surface immediately below the sphere. This is the distance between the $z = 0$ surface and the location of the sphere bottom. According to Hertzian contact theory [6], the maximum displacement is given by Eq. (1). Figure 12 shows this relationship with respect to applied force, as a dashed blue line. The experimental displacements, recorded by the camera, are represented by red circles, and the numerical simulation predictions are plotted as green triangles. From the figure, it is apparent that the experimental displacements follow the trend of the theory and numerical simulations, as expected: recall that the Young's Modulus is determined by fitting the theoretical and experimental displacements through Eqs. (1) and (2); hence this agreement between experiments, theory and numerics is expected.

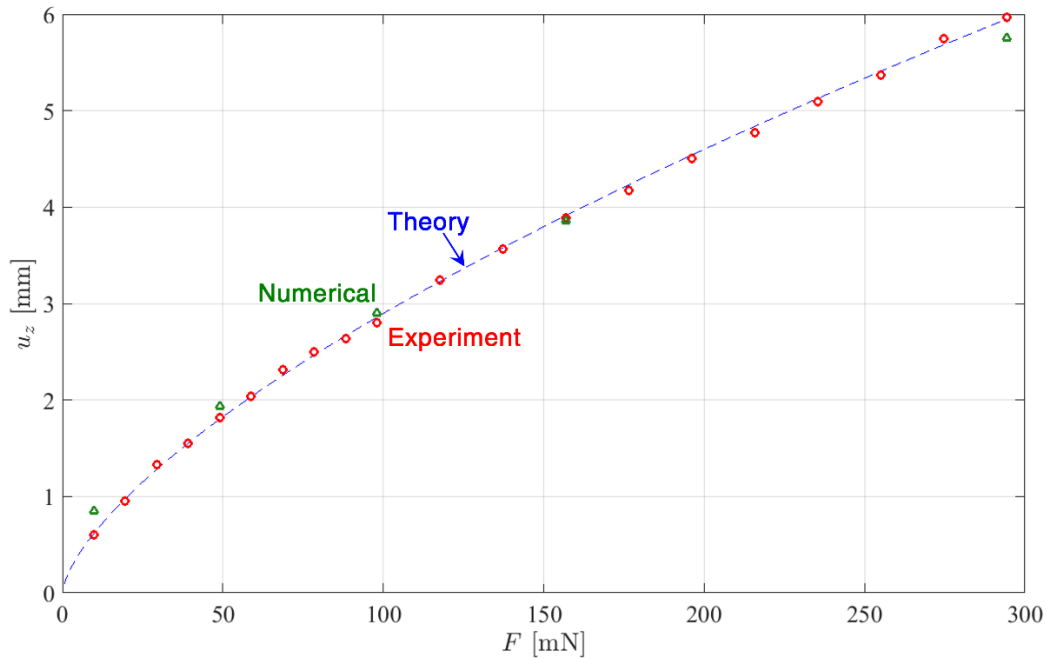

Figure 12: Relationship between maximum displacement and applied force for theory (blue dashed line, Eq. (1)), numerical (green triangle), and experiment (red circles).

The theoretical radius of contact between the surface of the gelatin and sphere is given by Eq. (8), where contact radius, $a$, scales with the applied force to the 1/3$^{rd}$ power. This relationship is shown in Fig. 13 as a dashed blue line, while the numerical contact radius is plotted with green triangles. The experimental contact radius is defined as the distance between the center of the sphere and the point where the sphere separates from the gelatin, which is determined using the camera. It is noted that the experimental uncertainty in contact radii measurements are larger than the uncertainty in determining the maximum displacement, due to the ambiguity in identifying the radial coordinate where the sphere separates from the gelatin. The experimental data points are represented by red circles in Fig. 13, where reasonable agreement is established with respect to theory and numerical simulations. The differences here are only a fraction of a millimeter. This agreement supports the assumptions made in Section 3 and appears to be valid even for the largest loading scenario.

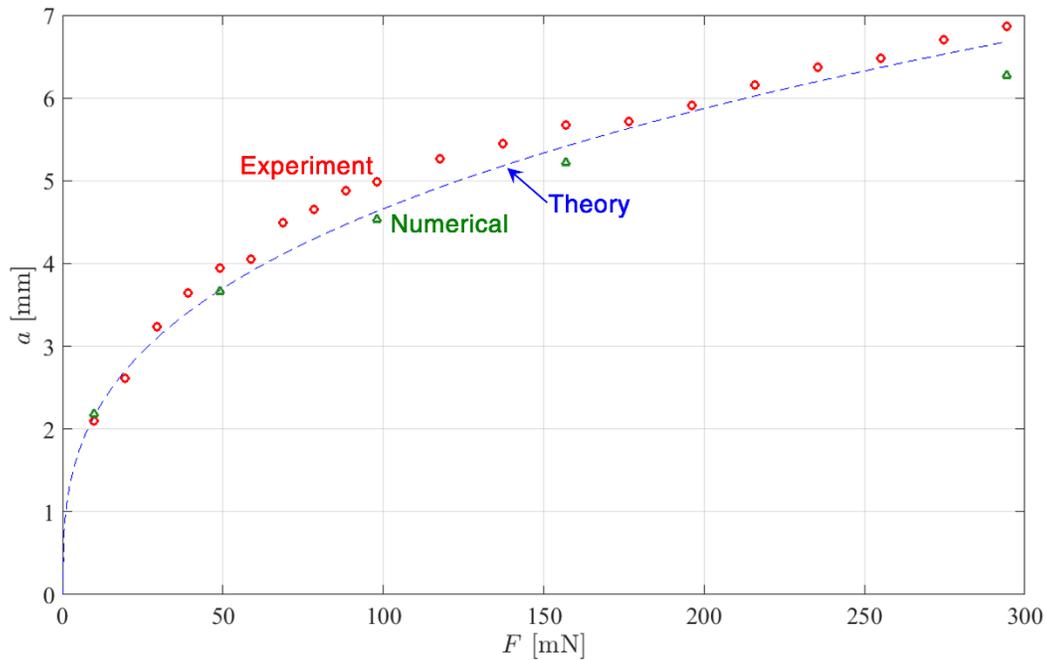

Figure 13: Relationship between contact radius and applied force for theory (blue dashed line, Eq. (8)), numerical (green triangle), and experiment (red circles).

# 6. Summary

Integrated photoelasticity is used to analyze the Hertzian contact problem of a rigid sphere loaded onto the top surface of a soft-solid, gelatin. The theoretical stress state of the gelatin is derived using the Love stress function and the Hankel transform method, which provides the stress tensor throughout the entire problem domain. The stress inside the gelatin is also calculated using numerical simulations. Both theoretical and simulated stress states are used in conjunction with the optically equivalent model to predict the phase retardation field expected in an integrated photoelasticity experiment. Experiments are carried out on a gelatin cuboid where excellent agreement is observed between theoretical and numerical predictions and the experimental phase retardation fields. A non-linear correlation is established between the maximum phase retardation and maximum equivalent stress for a variety of sphere loading conditions. This allows one to determine the maximum stress state, as well as stress components, in the Hertzian contact problem solely by conducting an integrated photoelasticity experiment. This is important because it identifies the position and magnitude where stress is maximum and, therefore, signifies where material yielding or failure is expected to occur first.

The agreement between experimental and theoretical results is also assessed though the maximum surface displacement and contact radius. The experiments show excellent agreement with theory according to surface displacement, which is expected as the Elastic Modulus was determined from experiments, while good agreement is observed regarding the contact radius. The successful determination of maximum equivalent stress and principal stress components at this location using integrated photoelasticity suggests that this method can be applied to similar, axisymmetric loading scenarios. Future studies will explore the viability of measuring equivalent stress and principal stress components for dynamic loads on gelatin media, such as the impact of a liquid droplet [41] or a jet [42]. Knowing the stress components during droplet impact and collapse will be essential in order to understand the associated material erosion mechanisms during such phenomena.

# LIST OF REFERENCES

induced by needle-free injection of a highly focused microjet. *Scientific reports*, *11*(1), 1-10.